\documentclass[12pt]{iopart}

\usepackage{graphicx}

\usepackage{todonotes,verbatim}
\usepackage{iopams}

\usepackage{hyperref}

\hypersetup{colorlinks=true,urlcolor=blue,linkcolor=blue,citecolor=blue}

\def\beq {\begin{equation}}
\def\eeq {\end{equation}}

\def\w {\omega}

\def\bfr {\mathbf{r}}
\def\bfq {\mathbf{q}}
\def\bfG {\mathbf{G}}

\begin{document}

\title{Local-field effects on the plasmon dispersion of two-dimensional transition metal dichalcogenides}

\author{Pierluigi Cudazzo$^{1,2}$, Matteo Gatti$^{3,2}$  and Angel Rubio$^{1,2,4}$}

\address{$^1$ Nano-Bio Spectroscopy group, Dpto. F\'isica de Materiales, Universidad del Pa\'is Vasco, Centro de F\'isica de Materiales CSIC-UPV/EHU-MPC and DIPC, Av. Tolosa 72, E-20018 San Sebasti\'an, Spain}
\address{$^2$ European Theoretical Spectroscopy Facility (ETSF)}
\address{$^3$ Laboratoire des Solides Irradi\'es, \'Ecole Polytechnique, CNRS-CEA/DSM,  F-91128 Palaiseau, France }
\address{$^4$ Fritz-Haber-Institut der Max-Planck-Gesellschaft, Theory Department, Faradayweg 4-6, D-14195 Berlin-Dahlem, Germany}

\ead{pierluigi.cudazzo@ehu.es,angel.rubio@ehu.es}

\date{\today}

\begin{abstract}

Two-dimensional transition-metal dichalcogenides (TMDs) are gaining
increasing attention as alternative to graphene for their very high
potential in optoelectronics applications. Here we
consider two prototypical metallic 2D TMDs, NbSe$_2$ and TaS$_2$.
 Using a first-principles approach, we investigate the properties of the 
 localised intraband $d$ plasmon that cannot be modelled on the basis of the
homogeneous electron gas. Finally, we discuss the effects of the reduced
dimensionality on the plasmon dispersion through the interplay between
interband transitions and local-field effects. This result can be
exploited to tune the plasmonic properties of these novel 2D materials.

\end{abstract}

\maketitle


\section{Introduction}

The isolation of graphene in 2004 \cite{Novoselov2004} gave a new twist to the field of two-dimensional (2D) crystals and boosted new avenues in condensed matter physics \cite{Novoselov2005,CastroNeto2009,Wang2012,Chhowalla2013,angel}. 
The unconventional electronic properties of graphene, that can be described in terms of massless 2D Dirac particles, have attracted the enthusiastic attention of the scientific community \cite{Novoselov2005b,Zhang2005} and promise a large variety of exciting technological applications \cite{geim2007}. 
However, graphene lacks a bandgap, preventing its use in electronic devices like transistors.
Hence finding the best way to induce a gap in graphene has been a very active research topic in the past few years. This can be obtained for example by chemically functionalisation as for graphane and other graphene derivatives \cite{Sofo2007,Elias2009,Robinson2010,Nair2010}, 
although at the price of loosing some of the unconventional properties of pristine graphene.  
To overcome these difficulties while keeping the advantages of the reduced dimensionality, the research activity is rapidly extending to other families of layered materials. In particular, transition-metal dichalcogenides (TMDs) \cite{Wilson1969} are increasingly gaining attention \cite{Coleman2011,Wang2012,Chhowalla2013}. Their chemical formula is MX$_2$, where M is a transition metal and X a chalcogen (S, Se,Te). In contrast to graphene, these materials show versatile electronic properties that vary from metallic to insulating, depending on the metal M \cite{Ataca2012}. 

TMDs can be shaped into monolayers, displaying distinct physical properties from their bulk counterpart. MoS$_2$, for example, has an indirect bandgap in bulk form, but in 2D the bandgap increases and becomes direct \cite{Splendiani2010,Mak2010}, being most favorable for optoelectronic applications. A 2D MoS$_2$ transistor has also recently realized \cite{Radisavljevic2011}, demonstrating that TMDs can be  used for flexible electronic devices. 
MoS$_2$ has shown to be promising also for spintronics applications \cite{Xiao2012}, absent in graphene due to its vanishing spin-orbit coupling, 
and in ``valleytronics'', exploiting the possibility of controlling the population of the valleys (i.e. conduction band minima) \cite{Mak2012,Zeng2012}.

Metallic TMDs such as NbSe$_2$ and TaS$_2$ exhibit remarkable low-temperature phenomena including the competition between superconductivity and charge-density wave (CDW) order \cite{Rossnagel2011}. 
In the single-layer form the hope is that they can be efficiently used for plasmonics applications with the aim of controlling light on the nanoscale \cite{Maier2007}. 
This typically requires surface plasmons that can confine the electromagnetic field at scales smaller than the wavelength of light. In fact, in comparison to conventional metals like gold or silver, 2D materials have a significant advantage: their plasmonic properties can be easily controlled by external means (e.g. doping or electrical voltage) \cite{Fei2012,Koppens2011,garciaabajo13}.

In view of these potentialities, it is very timely to address the study of the elementary electronic excitations of this novel class of 2D materials on the basis of predictive and reliable first-principles methods. Hence here we  make use of cutting-edge approaches based on time-dependent density-functional theory (TDDFT) \cite{onida02} to take into account the details of the electronic interactions at play. We calculate the dispersion (i.e. the dependence on the momentum transfer) of plasmons in two prototypical TMDs: NbSe$_2$ and TaS$_2$. To disclose the effects of the reduced dimensionality, we compare the plasmon properties in the bulk \cite{vanWezel11,Cudazzo12,Faraggi2012} and in the single-layer forms, 
showing the prominent role played by the induced microscopic electronic charges (the ``local fields'') in 2D.
This suggests the possibility to effectively tune the dielectric properties of these novel materials for plasmonics applications.

\section{Methods}

Plasmon excitations can be identified by the peaks in the loss function $L(\bfq,\w) = -\textrm{Im} \epsilon_M^{-1}(\bfq,\w)$, which
can be written in terms of the real and imaginary parts of the macroscopic dielectric function $\epsilon_M=\epsilon_1+i\epsilon_2$:
\beq
L(\bfq,\w)=-\textrm{Im} \frac{1}{\epsilon_M(\bfq,\w)}=\frac{\epsilon_2(\bfq,\w)}{[\epsilon_1(\bfq,\w)]^2+[\epsilon_2(\bfq,\w)]^2},
\label{losseps}
\eeq
where $\bfq$ is the momentum transfer. Plasmons energies $\w_p(\bfq)$ correspond to zeroes of $\epsilon_1(\bfq,\w_p)$ where the damping given by $\epsilon_2(\bfq,\w_p)$ is not too large so that $\epsilon_M\simeq 0$. In fact this represents the condition that must be satisfied by an electronic system  to sustain longitudinal collective excitations (i.e. the induced electric field is finite even in absence of external sources) \cite{grosso}. 
 
The macroscopic dielectric function $\epsilon_M$ can be calculated from first-principles, using methods based on Green's functions or within TDDFT \cite{onida02}, in which the Dyson-like equation for the linear-response polarizability $\chi$ reads:
\begin{eqnarray}
\chi(\bfr_1,\bfr_2,\w) = \chi_{KS}(\bfr_1,\bfr_2,\w)  \nonumber \\
+ \int d\bfr_3d\bfr_4 \,\, \chi_{KS}(\bfr_1,\bfr_3,\w) [v(\bfr_3,\bfr_4)+f_{xc}(\bfr_3,\bfr_4,\w)]
\chi(\bfr_4,\bfr_2,\w),
\label{dyson}
\end{eqnarray}
where $\chi_{KS}$ is the Kohn-Sham non-interacting polarizability that can be expressed in terms of the Kohn-Sham energies $\epsilon_{\mathbf{k}_ii}$ and orbitals $\phi_{\mathbf{k}_ii}(\mathbf{r})$ as:
\begin{eqnarray}\label{chiks1}
\chi_{KS}(\mathbf{r}_1,\mathbf{r}_2,\omega) &=& 2\sum_{\mathbf{k}_1i\mathbf{k}_jj}\phi_{\mathbf{k}_ii}(\mathbf{r}_1)\phi^*_{\mathbf{k}_jj}(\mathbf{r}_1)\phi^*_{\mathbf{k}_ii}(\mathbf{r}_2)\phi_{\mathbf{k}_jj}(\mathbf{r}_2) \nonumber \\
&\times& \left[\frac{\theta(\epsilon_{\mathbf{k}_ii}-\mu)\theta(\mu-\epsilon_{\mathbf{k}_jj})}{\omega-(\epsilon_{\mathbf{k}_ii}-\epsilon_{\mathbf{k}_jj})+i\eta}-\frac{\theta(\mu-\epsilon_{\mathbf{k}_ii})\theta(\epsilon_{\mathbf{k}_jj}-\mu)}{\omega-(\epsilon_{\mathbf{k}_ii}-\epsilon_{\mathbf{k}_jj})-i\eta}\right],
\end{eqnarray}
$v$ the Coulomb interaction and $f_{xc}$ the exchange-correlation TDDFT kernel.
In the calculations we adopted the local-density approximation (LDA) for $\chi_{KS}$ and the random-phase approximation (RPA) for $\chi$, which corresponds to setting $f_{xc}=0$ in Eq. \eref{dyson}. 
From $\chi$ one obtains the microscopic dielectric function $\epsilon{}$ since:
\beq
\epsilon^{-1}(\bfr_1,\bfr_2,\w)=\delta(\bfr_1-\bfr_2) + \int d\bfr_3 v(\bfr_1,\bfr_3)\chi(\bfr_3,\bfr_2,\w),
\label{eqloss}
\eeq
and the macroscopic dielectric function $\epsilon_M$ through \cite{adler62,wiser63}:
\beq
\epsilon_M(\bfq,\w) = \frac{1}{\epsilon^{-1}_{\bfG=0,\bfG'=0}(\bfq,\w)},
\eeq
where $\bfG$ and $\bfG'$ are reciprocal-lattice vectors and $\epsilon^{-1}_{\bfG,\bfG'}(\bfq,\w)$ is the Fourier transform to reciprocal space of $\epsilon^{-1}(\bfr_1,\bfr_2,\w)$ (with the momentum transfer $\bfq$ belonging to the first Brillouin zone).
The loss function $L(\bfq,\w)$ is thus:
\beq
-\textrm{Im} \epsilon^{-1}_M(\mathbf{q},\omega) = -v_0(\bfq)\chi_{0,0}(\bfq,\w).
\label{eqloss2}
\eeq
(where the $\bfG$ and $\bfG'$ indexes are understood).
In the homogeneous electron gas (HEG) the microscopic dielectric function $\epsilon$ is diagonal in $\bfG$ and $\bfG'$ and   $\epsilon_M(\bfq,\w) = \epsilon_{\bfG=0,\bfG'=0}(\bfq,\w)$.
In a system that is inhomogeneous and polarizable on the microscopic scale, instead, off-diagonal elements of $\epsilon$ are not zero and they all contribute 
to the head element ($\bfG=0,\bfG'=0$) of $\epsilon^{-1}$,  reflecting the so-called crystal local-field effects (LFE).

Alternatively, the macroscopic dielectric function can be directly calculated from \cite{onida02}:
\beq
\epsilon_M(\bfq,\w) = 1- v_0(\bfq)\bar{\chi}_{0,0}(\bfq,\w),
\label{chibarra}
\eeq
where the modified polarizability $\bar\chi$ satifies a Dyson equation like \eref{dyson} with a modified Coulomb interaction $\bar v$ 
that in the reciprocal space is equal to $v$ for all the $\bfG$ components but $\bfG=0$, for which $\bar v(\bfG=0)=0$:
\begin{eqnarray}
\bar \chi(\bfr_1,\bfr_2,\w) = \chi_{KS}(\bfr_1,\bfr_2,\w)  \nonumber \\
+ \int d\bfr_3d\bfr_4 \,\, \chi_{KS}(\bfr_1,\bfr_3,\w) [\bar v(\bfr_3,\bfr_4)+f_{xc}(\bfr_3,\bfr_4,\w)]
\bar \chi(\bfr_4,\bfr_2,\w).
\label{dyson2}
\end{eqnarray}
In the RPA with $f_{xc}=0$ setting also $\bar v=0$ in the Dyson equation \eref{dyson2}  means neglecting the LFE in the calculation of $\epsilon_M$, since in this case $\bar \chi = \chi_{KS}$.
Instead, with the inclusion of the LFE, $\chi_{KS}$ is screened by the short-range charge fluctuations associated to the induced Hartree potential.

Plasmon were first studied in the homogeneous electron gas \cite{Pines}, where the plasmon energy evaluated in RPA has a positive dispersion,
which is parabolic as a function of $q$ in 3D  and behaves as $\sqrt{q}$ in 2D \cite{vignale}.  The HEG model, which is characterized by a single parabolic band, is often taken as reference system also to study and interpret the collective excitations in real metallic materials. 
 However, in realistic materials deviations from the HEG RPA behavior can have different origins: 
 different electronic structure (i.e. non-parabolicity), effect of oscillator-strenght matrixelements, non-homogeneity, presence of interband transitions, correlation effects beyond RPA [i.e. due to $f_{xc}$ in Eq. (\ref{dyson})]. 
 In fact, results at variance with HEG, including  negative dispersions, have been reported in many cases, as for example in heavy alkali metal like Cs \cite{vomfelde89,karlsson94}, doped molecular crystals \cite{Cudazzo11} and 3D TMDs \cite{vanWezel11,Cudazzo12,Faraggi2012}.
 
Therefore, especially in view of the negative plasmon dispersion found in the bulk materials, 
it makes sense to address here in detail the investigation of plasmons in the 2D TMDs using a microscopic
approach based on a first-principles calculation of the realistic band structure beyond the HEG model.

\section{Computational details}

In our approach the dielectric function has been evaluated on top of LDA calculations implemented in a plane-wave-basis framework using norm-conserving Troullier-Martins pseudopotentials. 
We have used Abinit \cite{abinit} for the ground-state calculations and Yambo \cite{yambo} for the RPA spectra.
All the systems  have been simulated
using a two-dimensional triangular lattice with a basis of one MX$_2$ unit. 
We  have used the theoretical lattice constant (3.37 {\AA} for TaS$_2$ and 3.38 {\AA} for NbSe$_2$)
calculated using a $10\times 10\times 1$ $\mathbf{k}$ point mesh and an energy cutoff
for the plane-wave expansion between 30 and 40 Ha. Moreover, in the supercell approach, the distance between
adjacent layers has been set to 40 \AA. 
For the calculation of the dielectric function we have used a $30\times 30\times 1$ grid of $\mathbf{k}$ point with 80 band. 
Finally, local-field effects have been included inverting a matrix of rank 300 $\mathbf{G}$ vectors.

\section{Results and discussion}

\begin{figure}[ht]
 \centering
\includegraphics[width=.5\linewidth]{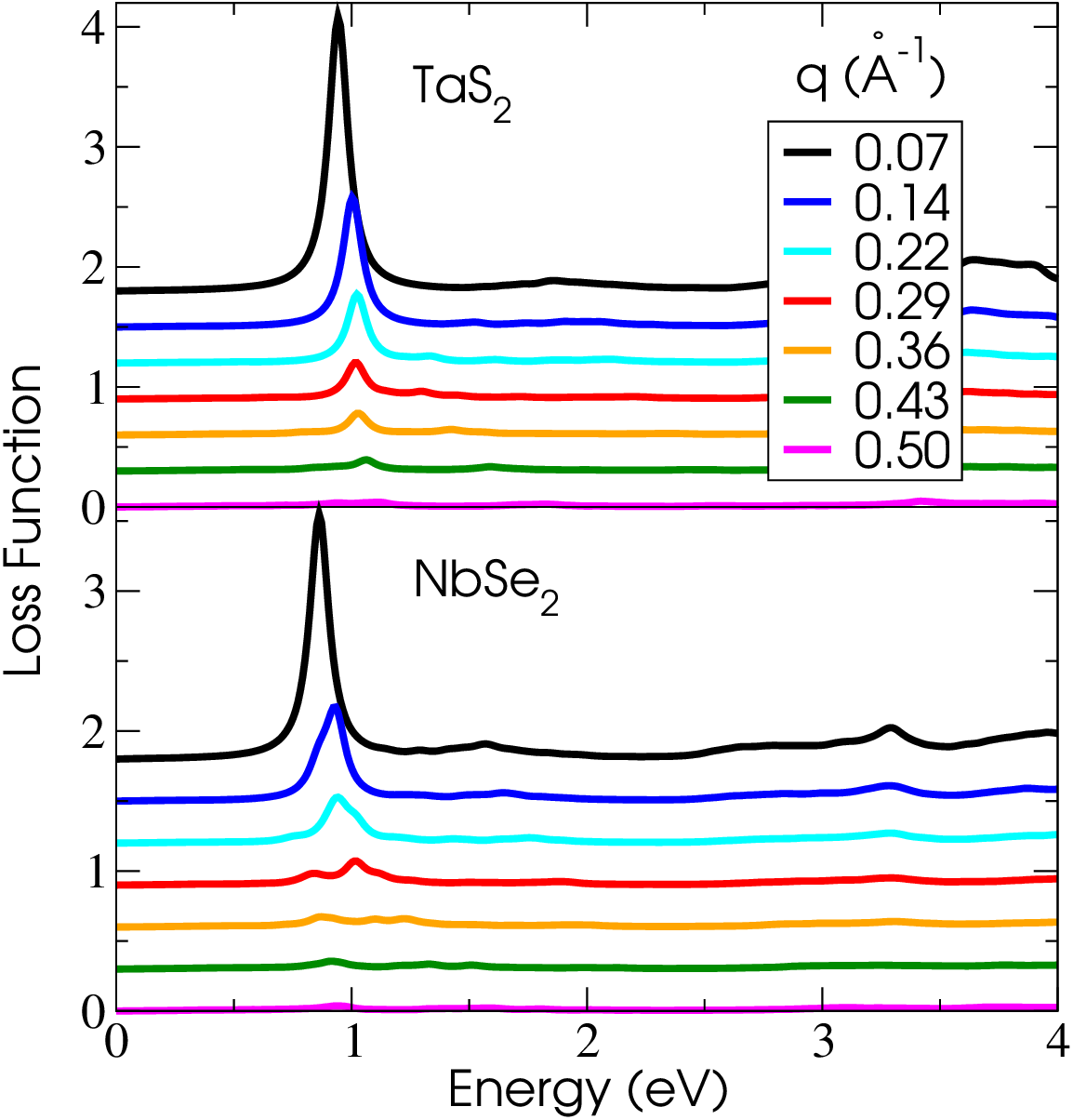}
\caption{Loss function of the single layers of  TaS$_2$ and NbSe$_2$ as a function of the momentum transfer $\mathbf{q}$ along the $\Gamma$M  direction. }
\label{fig1}
\end{figure}

\begin{figure}[t]
 \centering
\includegraphics[width=.45\linewidth]{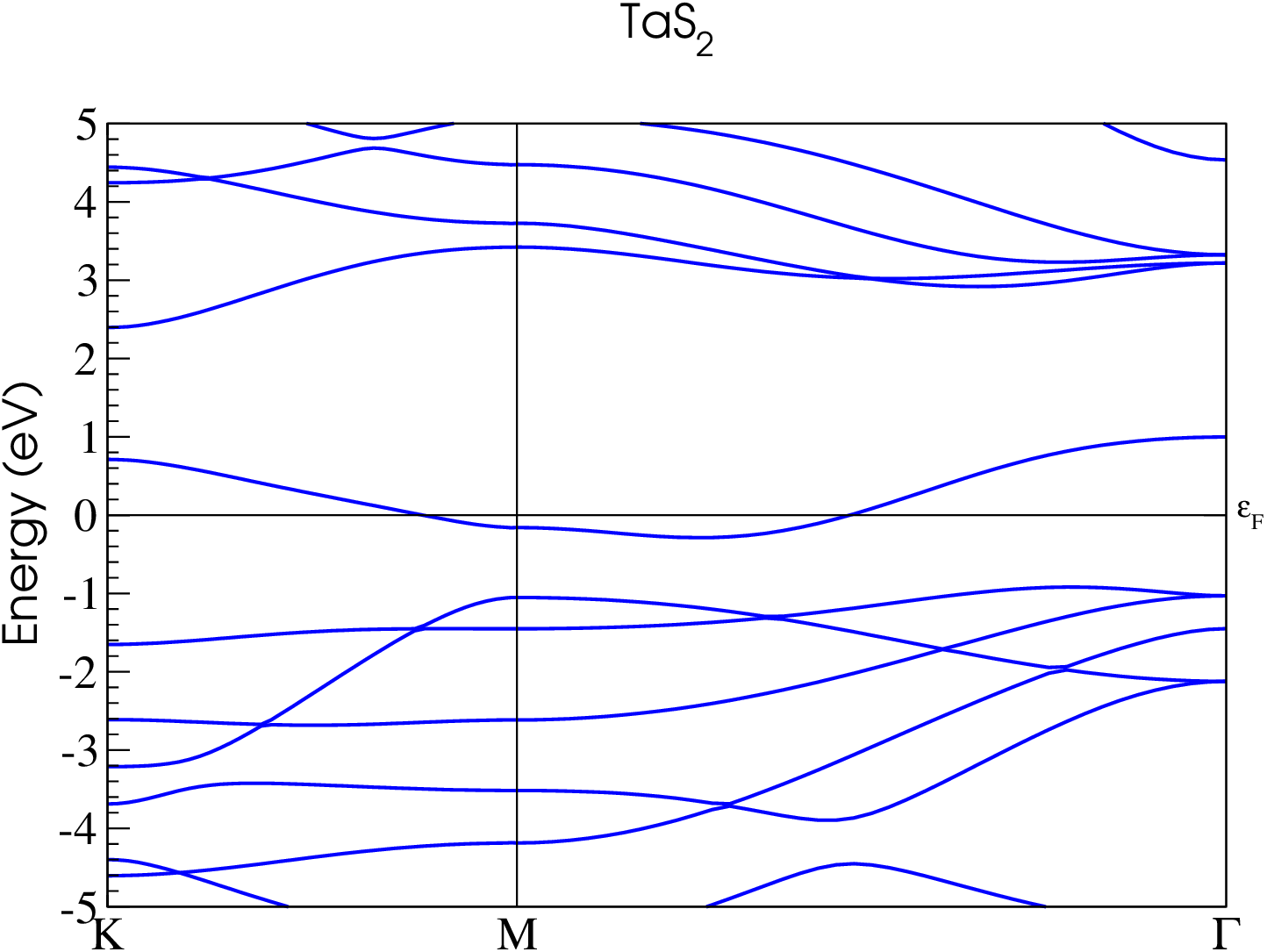}
\includegraphics[width=.45\linewidth]{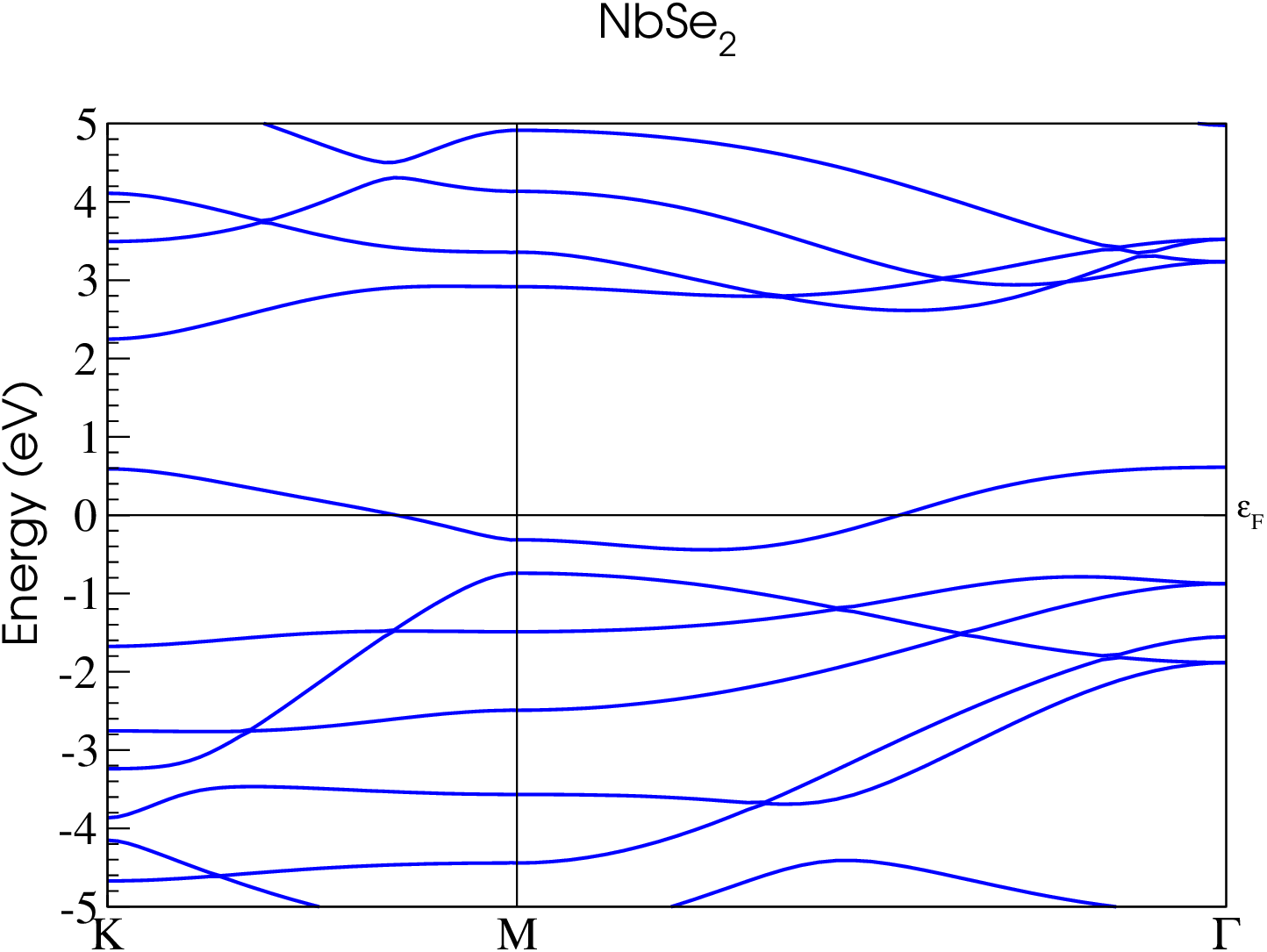}
\caption{LDA band structure of single-layer TaS$_2$ and NbSe$_2$.}
\label{fig1a}
\end{figure}

In Fig. \ref{fig1} we show the calculated low-energy RPA loss function of the single layers of  TaS$_2$ and NbSe$_2$ as a function of the momentum transfer $\mathbf{q}$ along the $\Gamma$M  direction\footnote{For the high-energy part of the spectrum at $\bfq=0$, comprising the $\pi$ and $\pi+\sigma$ plasmons, we refer to Ref. \protect\cite{Johari2011}. However, those calculations do not take into account LFE, which are negligible in the bulk, but are the dominant contribution in 2D, as we discuss in the following. On the other hand at $\bfq=0$ the intra-band plasmon is expected to have zero energy for dimentionality reasons.}. 
In the following we focus on TaS$_2$, but the same analysis holds also for NbSe$_2$, which has an electronic  band structure that is similar to TaS$_2$ (see Fig. \ref{fig1a}).
The main peak in the spectrum, which for TaS$_2$ at $\mathbf{q}=0.07$ \AA$^{-1}$ is located at $\omega=0.94$ eV 
is related to a zero of the real part of the dielectric function $\epsilon_1$ (see dashed green curve in Fig. \ref{fig2}). 
We thus identify this main feature as a plasmon resonance that derives from electron-hole pairs belonging to the $d$ band crossing the Fermi energy (see Fig. \ref{fig1a}).
In fact, these intraband transitions give rise to a peak of $\epsilon_2$  (at $\omega=0.70$ eV for this small $\bfq$, see solid green curve in Fig. \ref{fig2}) just below to the plasmon frequency.  
This implies that the plasmon is a collective excitation involving the charge density related to this metallic band. 
Therefore it has the same nature as the charge-carrier plasmon of the 3D bulk systems \cite{Cudazzo12,Faraggi2012}.
 
By increasing $\mathbf{q}$ we observe a positive dispersion of the plasmon peak, which looses intensity and becomes broader, entering the particle-hole continuum at $\mathbf{q}=0.35$ \AA$^{-1}$ (i.e. where the loss function gets to coincide to $\epsilon_2$). The plasmon bandwidth is rather small, only $\sim$ 0.1 eV, implying that the plasmon is quite localised in real space.
This positive dispersion, that is common to all the metallic single layers of the 2H family, 
is opposite to the bulk TMDs, where it has recently shown that the plasmon has a negative dispersion \cite{vanWezel11,Cudazzo12,Faraggi2012}. 
The similarity between the electronic band structure in 2D and 3D TMDs suggests 
that the opposite slope of the plasmon dispersion in the bulk and in the monolayer must be a direct consequence of the different dimensionality, as we are now going to show. 

\begin{figure}[t]
 \centering
\includegraphics[width=.55\linewidth]{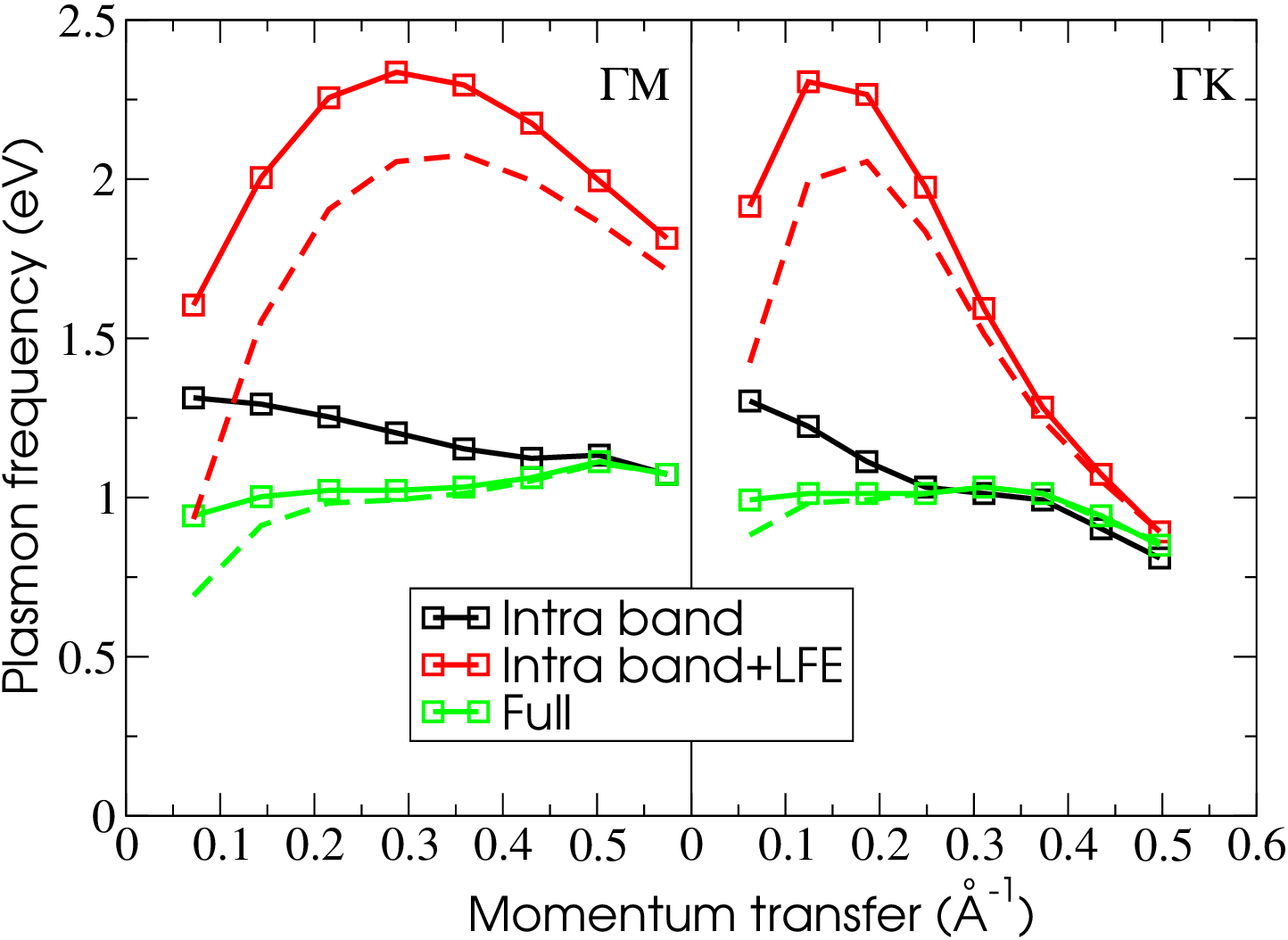}
\caption{Dispersion of the plasmon energy (solid lines) in TaS$_2$ as a function of $\bfq$ along $\Gamma$M and $\Gamma$K. The dashed lines correspond to the peak of the electron-hole continuum (see Fig. \protect\ref{fig2}). Different results are obtained taking into account only intraband transitions with or without local-field effects (red and black lines, respectively) and in the full calculation that includes also interband transitions (green line).}
\label{fig3}
\end{figure}

The non-interacting polarizability $\chi_{KS}$ can be split into the sum of an intraband contribution $\chi^{intra}_{KS}$ from the band crossing the Fermi level and an interband term $\chi^{inter}_{KS}$ from all the other electron-hole transitions: $\chi_{KS}=\chi^{intra}_{KS}+\chi^{inter}_{KS}$.
In the first stage, we calculate the ``bare plasmon'' considering only intraband transitions  and neglecting the LFE.
The plasmon is thus determined only by the details of the band structure and the matrix elements of the oscillator strenghts entering $\chi^{intra}_{KS}$.
As in the bulk,  and in contrast to the HEG, we find that the intrinsic dispersion of the bare plasmon is negative (see solid black line in Fig. \ref{fig3}). This is a consequence of the presence of non-dispersive $d$ bands at the Fermi level both in 3D and 2D \cite{Cudazzo12}, which makes these materials very different from the HEG, as we now discuss more in details for the 2D case.

\begin{figure}[t]
 \centering
\includegraphics[width=.55\linewidth]{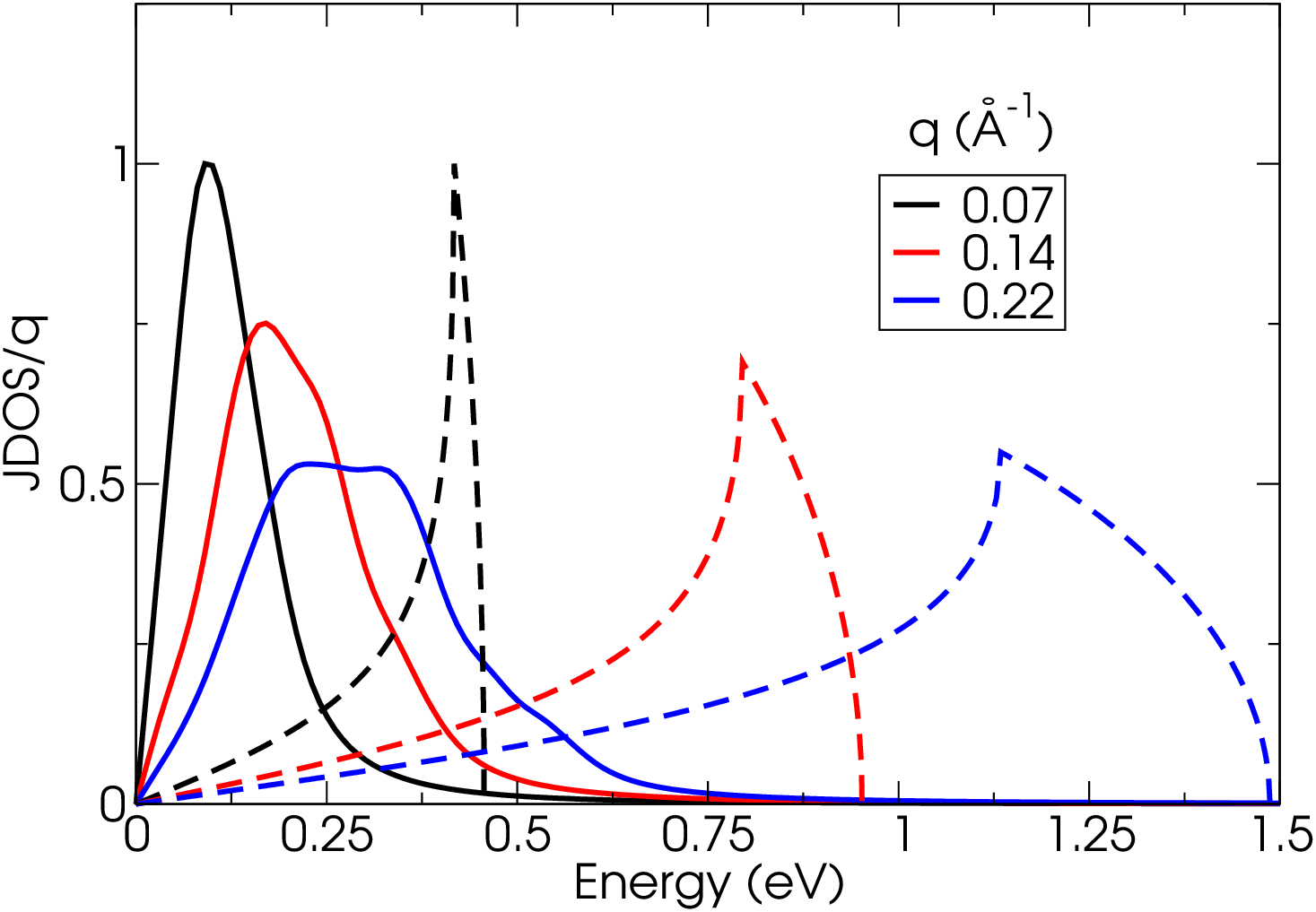}
\caption{Joint density of states JDOS$(\bfq,\w)/|\bfq|$ for TaS$_2$ (solid lines) from intraband transitions involving the $d$ band and for an equivalent electron number for the 2D HEG ($r_s=3.34$) (dashed lines). Both are normalized to the intensity of the respective highest peak. The JDOS of the HEG is dispersing with q, while for TaS$_2$ it is not.}
\label{jdos}
\end{figure}

\begin{figure}[t]
 \centering
\includegraphics[width=.55\linewidth]{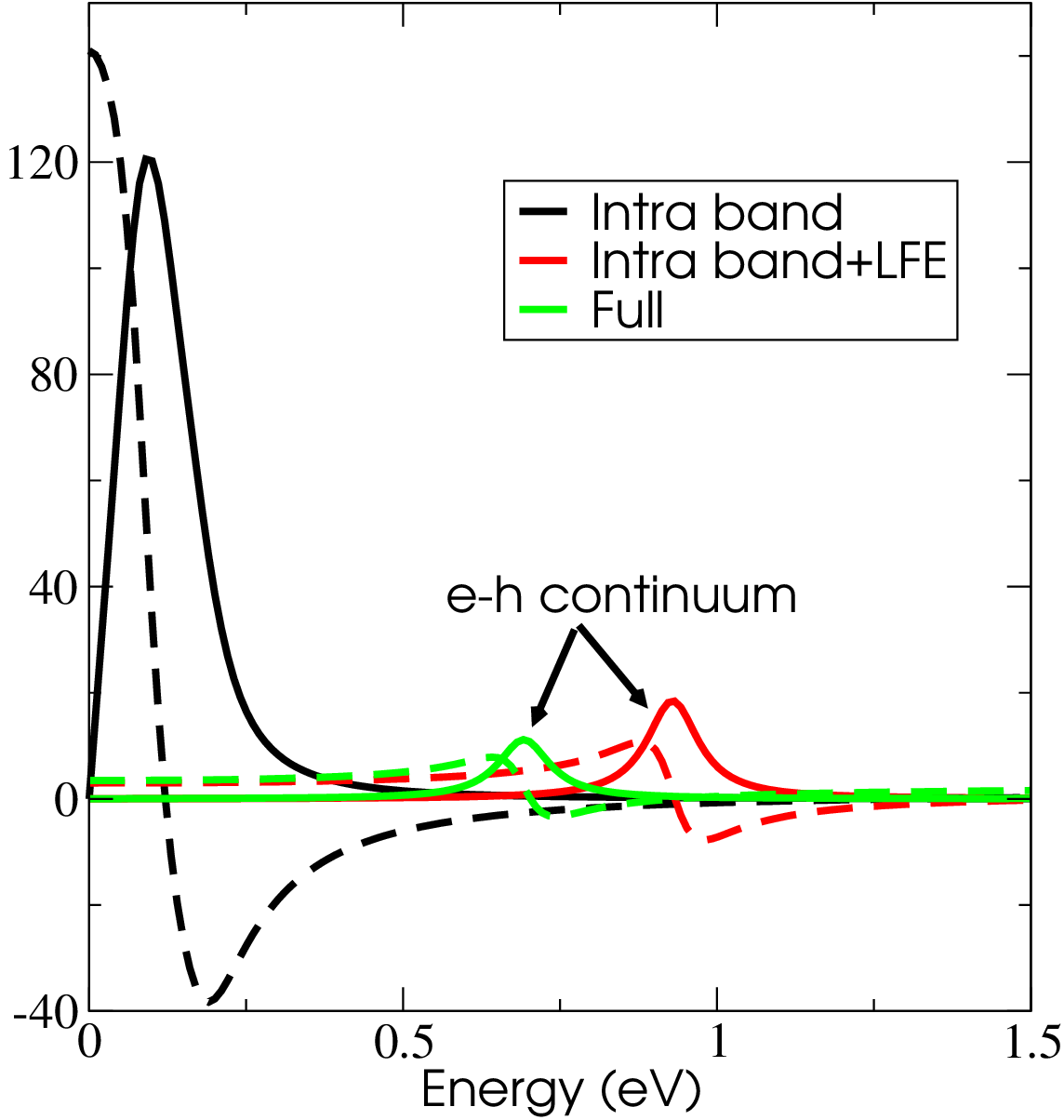}
\caption{Imaginary and real parts of the macroscopic dielectric function (solid and dashed lines, respectively) for $\bfq=0.07 \AA^{-1}$ along $\Gamma$M in TaS$_2$. The same approximations as in Fig. \protect\ref{fig3} are adopted.}
\label{fig2}
\end{figure}

In the HEG the positive dispersion of the plasmon frequency can be simply understood in terms of its band structure. 
The presence of a wide parabolic energy band gives rise to a sharp peak in the  joint density of states (JODS) with strong positive dispersion with $q$ (see dashed lines Fig. \ref{jdos}). Since in the 2D HEG  $\epsilon_2$ is proportional to the JDOS, $\epsilon_2(\bfq,\w)\propto JDOS(\bfq,\w)/|\bfq|$, 
this behavior is reflected, through the Kramers-Kronig relations, on the real part of $\epsilon$ and thus on the plasmon causing the $\sqrt{q}$ dependence.
In the real 2D TMDs, in contrast to the 2D HEG, the presence of a narrow $d$ band, gives rise to an intense and non-dispersive peak in the JDOS (see solid lines in  Fig. \ref{jdos}). Since the numerator of $\chi_{KS}^{intra}$ [see Eq. (\ref{chiks1})] does not play an essential role in these materials, this results in a negative bare plasmon dispersion (see solid black line in Fig. \ref{fig3}), as it has been shown  for the  3D TMDs in Ref. \cite{Cudazzo12}.
Therefore we conclude that, even though the HEG is often taken as reference system to describe collective excitations in metals, 
a picture based on the HEG model is not valid for the TMDs either in 3D or in 2D. In fact, also in the full calculation the plasmon dispersion does not follow the behaviour of the 2D HEG in the range of $q$ studied. 

The main difference between the bulk and the single layer is given by the effect of the crystal local fields. 
In the bulk TMDs LFE do not affect the loss function for in-plane momentum transfers: the plasmon dispersion remains negative also when LFE are taken into account.
On the contrary, LFE play a key role in the single layers that are characterized by an intrinsic strong inhomogeneity. LFE are the result of the mixing of different electron-hole transitions. 
We analyse them in two steps: first taking into account only the intraband contribution and then including the effect of the interband transitions.
Therefore in  Fig. \ref{fig2}  we compare the imaginary part of the dielectric function $\epsilon_2$ (see solid lines) in three different approximations: 
(i) including only intraband transitions without LFE (black lines); 
(ii) including only intraband transitions with LFE (red lines); 
(iii) the full calculation adding also interband transitions (green lines). 
We see that the main effect of local fields on $\epsilon_2$ derive from the intraband transitions (compare black and red lines in Fig. \ref{fig2}). 
It consists into a blueshift (by about 0.8 eV) and a strong damping (by more than 80$\%$) of the peak of $\epsilon_2$. 
This depolarization effect on $\epsilon_2$ is due to the repulsive nature of $\bar v$ in Eq. \eref{dyson2}.
From the expression of $\bar{\chi}$ in Eq. \eref{dyson2} we can also understand that LFE are important when $\chi_{KS}$ is large (i.e. the system is strongly polarizable) and 
when the electronic wavefunctions are strongly localized and the system is inhomogeneous.
In fact, a similar depolarization effect has been observed for example also for the $\pi$ plasmon in graphene \cite{hambach08}.
In 2D TMDs we find an even larger effect of local fields. While graphene is characterized by a zero bandgap and $\pi$ states, 
the metallic band in 2D TMD is related to $d$ states that are more strongly localized.
These depolarization effects on $\epsilon_2$ reflect also on $\epsilon_1$ through the Kramers-Kronig relations (see dashed lines in Fig. \ref{fig2}). 
In fact, due to the suppression of $\epsilon_2$, the zero of $\epsilon_1$ that sets the bare plasmon energy gets closer to the continuum of electron-hole transitions (i.e. the peak in $\epsilon_2$). 

Thus when LFE are taken into account the plasmon is blueshifted (see Fig. \ref{fig3}) and damped by the independent-particle motion.
LFE cause a strong positive dispersion of the intraband peak in $\epsilon_2$ at small $\mathbf{q}$ (see red dashed line in Fig \ref{fig3}). 
Moreover, due to the strong depolarization, the plasmon follows the behavior of the electron-hole continuum (compare dashed and solid red lines in Fig. \ref{fig3}). 
In fact, in a localised system, where LFE are very large, the long-range part of the Coulomb interaction, which is the difference between $v$ and $\bar v$, becomes negligible and $\bar \chi$ and $\chi$ becomes very similar. In turn, since $\chi$ gives the loss function $-\textrm{Im} \epsilon_M^{-1}$ [see Eq. \eref{eqloss2}] and $\bar \chi$ gives $\textrm{Im} \epsilon_M$ [see Eq. \eref{chibarra}], this explains why at small $\mathbf{q}$ the plasmon follows the electron-hole continuum and then becomes completely damped.

We now consider also the contribution of interband transitions by adding  $\chi^{inter}_{KS}$ to $\chi^{intra}_{KS}$ and solving the Dyson equation for $\chi$ to include LFE. 
As we can see in Fig. \ref{fig2} (green line),  the effect of interband transitions is a redshift of the continuum of electron-hole transitions. 
As a consequence, also the zero of $\epsilon_1$ and hence the plasmon frequency results redshifted by about 0.7 eV. 
Moreover, the positive plasmon dispersion, found taking into account only intraband transitions and LFE, 
is strongly reduced (see green line in Fig. \ref{fig3}) with an increase of the damping. 

To better understand the interplay between interband transitions and LFE on the plasmon dispersion, it is useful to
introduce the intraband polarizability $\chi^{intra}$, which, dropping for the moment the spatial indexes for simplicity, is defined as:
\begin{equation}
\chi^{intra}(\w)=\chi^{intra}_{KS}(\w)+\chi^{intra}_{KS}(\w)W_{inter}(\w)\chi^{intra}(\w),
\label{intra}
\end{equation}
where $W_{inter}(\w)=\epsilon_{inter}^{-1}(\w)v$ is the Coulomb interaction that is dynamically screened by the interband transitions. 
Then the total polarizability $\chi$ can be written as:
\begin{equation}
\chi(\w)=\epsilon_{inter}^{-1}(\w) \{\chi^{intra}(\w)+[1-\chi^{intra}_{KS}(\w)W_{inter}(\w)]^{-1}\chi^{inter}_{KS}(\w)\}.
\label{eqchi}
\end{equation}
In absence of interband transitions, $\chi^{inter}_{KS}=0$ and $\epsilon_{inter}=1$. Thus  
$\chi$ reduces to the intraband response function $\chi^{intra}$ evaluated with the unscreened Coulomb potential, since $W_{inter}=v$.
At frequencies around the plasmon energy, $\chi^{inter}_{KS}$ is still very small and Eq. \eref{eqchi} simplifies to:
\beq
\chi(\w) \simeq \epsilon_{inter}^{-1}(\w) \chi^{intra}(\w)
\eeq
In this energy range, where $\epsilon_{inter}$ is real, we thus obtain for the loss function the following expression [see Eq. \eref{eqloss2}]:
\begin{eqnarray}
-\textrm{Im} \epsilon^{-1}_M(\mathbf{q},\omega) &=& -v_0(\bfq) [\epsilon_{inter}(\mathbf{q},\omega)]_{00}^{-1}
\textrm{Im}\chi^{intra}_{00}(\mathbf{q},\omega) \nonumber  \\
&&-v_0(\bfq) \sum_{\mathbf{G}\neq 0}[\epsilon_{inter}(\mathbf{q},\omega)]_{0\mathbf{G}}^{-1}\textrm{Im}\chi^{intra}_{\mathbf{G}0}(\mathbf{q},\omega).
\end{eqnarray}
By retaining only the zero-order contribution of the off-diagonal terms of $\chi^{intra}$ from Eq. \eref{intra}, we finally arrive at the following expresion for the loss functions:
\begin{eqnarray}
-\textrm{Im} \epsilon^{-1}_M(\mathbf{q},\omega) &=& -v_0(\bfq) [\epsilon_{inter}(\mathbf{q},\omega)]_{00}^{-1}
\textrm{Im}\chi^{intra}_{00}(\mathbf{q},\omega) \nonumber  \\
&&-v_0(\bfq) \sum_{\mathbf{G}\neq 0}[\epsilon_{inter}(\mathbf{q},\omega)]_{0\mathbf{G}}^{-1}[\textrm{Im}\chi^{intra}_{KS}]_{\mathbf{G}0}(\mathbf{q},\omega) + \ldots
\label{epsintra}
\end{eqnarray}
 From Eq. \eref{epsintra} it is thus clear that interband transitions have two effects. 
 First interband transitions screen  $\chi^{intra}$ causing a redshift of the plasmon and the continuum of the electron-hole transitions. 
 This is the effect of the first term in Eq. \eref{epsintra}.
 The second term in Eq. \eref{epsintra} is instead related to the LFE: it 
 is responsible for summing the off-diagonal part of $\chi^{intra}_{KS}$ dominated by single-particle excitations to the diagonal part of $\chi^{intra}$ dominated by collective excitations. This term causes a redshift of the spectrum and makes  the plasmon damping stronger.
 
 Moreover, as $\mathbf{q}$ increases, the screening from interband transitions becomes weaker, reducing its redshift effect on the plasmon energy. 
 As a result, the positive dispersion of the plasmon becomes stronger. 
 However, at the same time, as $\mathbf{q}$ increases, LFE contained in the second term of Eq. \eref{epsintra} become more and more important, ehnancing the coupling between the plasmon and the independent electron-hole transitions at lower energy. In turn, this tends to reduce the positive dispersion of the plasmon.
 We thus see that the interplay between interband transitions and LFE induces two competing effects on the dispersion of the intraband plasmon that manifest differently in 2D and 3D structures. 

 In the bulk 3D TMDs the LFE are negligible and the first term of Eq. \eref{epsintra} is always dominant. 
 As a consequence, the resulting effect of the interband transitions is to temper the negative dispersion of the bulk intraband plasmon. 
 On the other hand, in 2D systems, due to the large LFE, the second term in Eq. \eref{epsintra} becomes the most prominent contribution. 
 In this case, interband transitions reduce the positive dispersion of the intraband plasmon (compare red and green lines in Fig. \ref{fig3}).

\section{Conclusions}
 
 In summary, we have calculated the charge-carrier plasmon dispersion in two prototypical single-layer 
 metallic transition metal dichalcogenides (TMD): TaS$_2$ and NbSe$_2$.
 As in the bulk, the nondispersive metallic band prevents using the homogeneous electron gas as a model to describe the plasmon excitations in these materials.  
 Our analysis has shown that the plasmon dispersion of the 2D TMDs can be even more easily tuned than their 3D counterparts.
 In the bulk the plasmon dispersion is mainly determined by the joint density of the states crossing the Fermi level. The negative dispersion in the bulk can be switched to positive by doping with electrons or holes \cite{Cudazzo12}.
 In the monolayers, instead, the plasmon character is also the result of the interplay between local-field effects and interband transitions, 
 which gives rise to a small bandwidth corresponding to a localised plasmon. 
 By acting also on the interband transitions it is in principle possible to tune the plasmon nature through the microscopic charge response 
 that constitute the strong ``local fields'' acting on these materials. 
 Therefore, metallic 2D TMDs seem a promising platform to investigate possible applications in nanoplasmonics and demand a deeper consideration.

\ack

 We acknowledge financial support from the
European Research Council Advanced Grant
DYNamo (ERC-2010-AdG-267374), Spanish Grants (2010-21282-C02-01  and PIB2010US-00652),
Grupos Consolidados UPV/EHU del Gobierno Vasco (IT578-13) and European
Commission projects CRONOS (Grant number 280879-2 CRONOS CP-FP7).
Computational time was granted by  BSC Red Espanola de Supercomputacion and GENCI (Project number 544)

\section*{References}

 \bibliography{biblio}
 \bibliographystyle{iopart-num}

\end{document}